\documentclass[12pt]{article}
\usepackage{epsfig}

\newcommand{\cor}[1]{\left\langle{#1}\right\rangle}
\newcommand{\f}[2]{\frac{#1}{#2}}
\newcommand{\ra}{\longrightarrow}
\newcommand{\eq}{\begin{equation}}
\newcommand{\eqx}{\end{equation}}
\newcommand{\eqn}{\begin{eqnarray}}
\newcommand{\eqnx}{\end{eqnarray}}
\newcommand{\qqqq}{\quad\quad\quad\quad}

\newcommand{\al}{\alpha}

\title{{\bf IMPROVED INTERMITTENCY ANALYSIS OF SINGLE EVENT DATA}
       \thanks{e-mail: {\tt ufrjanik@jetta.if.uj.edu.pl},
                       {\tt beataz@qcd.ifj.edu.pl}}}
\author{Romuald A. Janik $^{\dag)}$ and
       Beata Ziaja $^{\dag\dag)}$\\
        $^{\dag}$ \it Institute of  Physics, Jagellonian University\\
        \it Reymonta 4, 30-059 Cracow, Poland\\
        \\
        $^{\dag\dag}$ \it Department of Theoretical Physics\\
        \it Institute of Nuclear Physics\\
        \it Radzikowskiego 152, 31-142 Cracow, Poland\\}
\date{May 1998}

\begin{document}

%
%
%
%

\maketitle

\begin{abstract}
The intermittency analysis of single event data (particle moments)
in multiparticle production is improved, taking into account corrections
due to the reconstruction of history of a particle cascade. This
approach is tested within the framework of the $\alpha$-model.
\end{abstract}

\newpage
\section{Introduction}

 The first data on possibile intermittent behaviour in multiparticle
production \cite{l1} came from the analysis of the single event of high multiplicity
recorded by the JACEE collaboration \cite{l2}. It was soon realized, however,
that the idea may be applied to events of any multiplicity provided that
averaging of the distributions is performed \cite{l3}. This led to many
successful experimental studies of intermittency \cite{l4}, and allowed to
express the effect in terms of the multiparticle correlation functions \cite{l5}.
It should be realized, however, that the averaging procedure, apart from clear
advantages, brings also a danger of overlooking some interesting effects
if they are present only in a part of events produced in high-energy collisions.
For example, the unique properties due to the
presence of quark-gluon plasma in multiparticle production would manifest
only in some events, see e.g. \cite{l6}. Taking into account the sample of events and
averaging over them destroys such an information. Therefore, as already discussed
in \cite{hwa}, \cite{bz},
there is a need for event-by-event analysis of multiparticle production
processes. In this way the fluctuations of the measured physical
quantities (e.\ g.\
factorial moments) from event to event can be observed and estimated,
and any anomalous behaviour of them has a chance to manifest very clearly.
Such studies should necessarily be restricted to high-multiplicity events because only
there one may expect the statistical fluctuations to be under control.

Such an approach to the multiparticle data analysis has been already
proposed in \cite{bsz}, \cite{hwa}, \cite{bz}. In \cite{hwa} a new
quantity: erraticity has been introduced to investigate the event-by-event
fluctuations of factorial moments, and to search for their properties.
Erraticity denotes the normalized moment of event-by-event distribution
of a horizontally averaged factorial moment. It probes both types
of fluctuations: horizontal ones connected with the spatial bin pattern and
vertical ones i.e. event-by-event ones.

In \cite{bz} the event-by-event fluctations of
particle moments have been investigated directly for the one-dimensional
$\alpha-$model of random cascading. Monte-Carlo simulations of the model
allowed one
to obtain the histograms of event-by-event distributions of horizontally
averaged particle momenta and estimate the relation between the intermittency
parameters obtained from such a histogram, and the intermittency parameters
derived after usual procedure of averaging particle moments over all events.
The results were promising: the average value of the intermittency exponent
reproduced well the value obtained by averaging particle moments over events,
however with the tendency to underestimate the theoretical value. Furthermore,
the dispersion of the moment distribution was inversely proportional to the
length of a generated cascade, and even for short cascades substantially smaller
than the average value. The latter property was of a special importance~:
it allowed one to distinguish between groups of events emerging from cascades
with different characteristics.

In this paper we would like to improve the analysis of single event data
presented in \cite{bz}. Taking into account corrections due to the method of
recovering the history of the multiparticle cascade \cite{l1}, \cite{l2}, we
expect to reduce the discrepancy between the theoretical value of intermittency
exponent and its value estimated from the event-by-event histogram \cite{bz}.
Our discussion will proceed as follows. In section 2 we recall the definition
of the intermittency exponents and the technique used to calculate them
\cite{l1}, \cite{l2}. In section 3 the definition of the $\al$ model
will be briefly presented, and applied in section 4 to calculate
corrections for extracting intermittency exponents from single event data.
Section 5 is devoted to the comparison of theoretical results with
numerical simulations. Finally in section 6 we present our conclusions.

\section{Intermittency exponents}

Consider a multiparticle production cascade distributed into $M$ bins.
At the $n^{th}$ stage of the cascade we measure the distribution of
a particle density into $M$ bins.
Assume for simplicity that $M=2^{n}$. We thus have $2^n$ numbers (quantities)
denoting the content of each bin~:
\eq
x^{(n)}_i \qqqq, i=0,1,\ldots,2^n-1.
\eqx
To perform the event-by-event analysis one is interested in the behaviour of
particle moments with the stage of the cascade~:
\eq
z^{(n)}_q=\f{1}{2^n} \sum_{i=0}^{2^n-1} \left( x^{(n)}_i \right)^q.
\eqx
The scaling behaviour of these moments is parametrized by intermittency
exponents $\phi_q$ \cite{l1}~:
\eq
z^{(n)}_q \sim 2^{n\cdot \phi_q}.
\label{e.intex}
\eqx
The task is to estimate the value of an intermittency exponent.
There are two different ways of doing it.
The first one is to calculate the average moment $z^{(n)}_q$
for the whole ensemble of individual events, and from this to reconstruct
the intermittency exponent.
The second one is to calculate the exponent $\phi_q$ for each
event separately, and then to recover the average $\phi_q$. The latter
approach has the advantage of being able to distinguish between two
independent cascading processes each with different $\phi_q$. This
could be done by looking at the distribution of individual $\phi_q$'s.
In the former method both of these possibly independent processes
would be artificially forced to be described by a single `effective'
$\phi_q$.

In the following we would like to address the question of reliably
reconstructing the correct value of $\phi_q$ from single event
data. Numerical simulations in \cite{bz} showed that there is an
inherent discrepancy between the theoretical value
and the distributions of event-by-event $\phi_q$ (see
Tables 1,2). The aim of this letter is to analyze this result
and introduce a correction which improves the estimation.

A convenient way of calculating $\phi_q$ is to make a linear fit to
the points $(n,\log z^{(n)}_q)$ ( all logarithms are taken to be
calculated in base 2, i.e. $\log x \equiv {\rm ln }x/{\rm ln 2}$)~:
\eq
\label{e.fit}
\log z^{(n)}_q = n\cdot \phi_q + b.
\eqx
This procedure has the advantage of cancelling out the major part of
the correction coming from the fact that we are effectively
reconstructing the exponents from $\cor{\log z^{(n)}_q}$ while the
true value is defined in terms of $\log\cor{z^{(n)}_q}$.

However there is still one caveat to (\ref{e.fit}). Since we cannot in
general separate out the various stages of the cascade, one
reconstructs the previous stages from the last one by summing the
$x^{(n)}_i$'s in adjacent bins using the technique described in \cite{l1}
( and applied there to JACEE event \cite{l2} ). Namely one approximates
the true value of $x^{(n-k)}_i$ by~:
\eq
x^{(n-k)}_i \ra y^{(n-k)}_i= \f{1}{2^k} \sum_{j=0}^{2^k-1}
x^{(n)}_{2^k\times i+j}.
\eqx
Therefore in (\ref{e.fit}) one really uses the reconstructed moments~:
\eq
z^{(k)}_{q,reconstructed}=\f{1}{2^k} \sum_{j=0}^{2^k-1}  \left(
y^{(k)}_i \right)^q.
\eqx

We will now use the $\al$ model of random cascading \cite{l1} to calculate
explicitly the difference between the true and reconstructed moments
and the resulting shift of the $\phi_q$ distribution from the theoretical
value.

\section{The $\al$ model of random cascading}

In the $\al$ model of random cascading \cite{l1} the root of the cascade
--- $x^{(0)}_0$
is taken to be $a$ with probability $p_a$ and $b$ otherwise (with
probability $p_b=1-p_a$). One generates the next stages of the cascade
recursively. The two bins $x^{(n+1)}_{2i}$ and $x^{(n+1)}_{2i+1}$ are
obtained from $x^{(n)}_i$ by~:
\eqn
x^{(n+1)}_{2i} \ra a \cdot x^{(n)}_i \qqqq \mbox{with probability $p_a$},\\
x^{(n+1)}_{2i} \ra b \cdot x^{(n)}_i \qqqq \mbox{with probability $p_b$},
\eqnx
and same for $x^{(n+1)}_{2i+1}$.
The parameters $a$ and $b$ are taken to satisfy~:
\eq
\label{e.norm}
ap_a+bp_b=1.
\eqx
Particle moments fulfill the relation~:
\eq
z^{(n)}_q=2^{(n+1)\cdot \phi_q},
\eqx
where intermittency exponents $\phi_q$ are equal to~:
\eq
\phi_q=\log (a^{q}p_a+b^{q}p_b).
\eqx

\section{Reconstructed moments}

The reconstructed moments in the $\al$ model are related to the true
ones by~:
\eq
z^{(n-k)}_{q,reconstructed}=
\f{1}{2^n}\sum_{i=0}^{2^{n-k}-1}
\cor{ \left(
 \sum_{j=0}^{2^k-1} x^{(n)}_{2^k i+j}\right)^q } \equiv z^{(n-k)}_q
\cdot p_q(k).
\eqx
where the average $\cor{\ldots}$ is taken over the random choices made
{\em only} above the $(n-k)$-th stage of the cascade.
The factor $p_q(k)$ can be calculated exactly (see below) and we
propose to use it to compensate for the errors introduced by the
reconstruction procedure. In particular the reconstructed moments
entering (\ref{e.fit}) will be shifted by~:
\eq
\log z^{(n-k)}_{q,reconstructed} \ra  \log z^{(n-k)}_{q,reconstructed}
 -\log(p_q(k)).
\label{imp}
\eqx


We will now determine the explicit form of the correction $p_q(k)$. By
the definition of the $\al$ model, the correction $p_q(k)$ can be
calculated just by evaluating~:
\eq
\label{e.pqk}
p_q(k)=\cor{\left( \f{1}{2^k}\sum_{i=0}^{2^k-1} x^{(k)}_i \right)^q}
\eqx
in the $\al$ model modified by taking the starting bin $x^{(0)}_0=1$.

First it is easy to see that for $q=1$ there is no correction
$p_1(k)=1$. This is due to (\ref{e.norm}). Also all corrections vanish
for $k=0$~:
\eq
\label{e.init}
p_q(0)=1.
\eqx
The appearance of a correction for $q>1$ comes from the fact that the
`number' of particles in this model has a nonzero dispersion.

Consider first the case of $q=2$.
We will now split the bins ($x_i$'s) appearing in (\ref{e.pqk}) into a
left half ($i<2^{k-1}$) and a right half ($i\geq 2^{k-1}$):
\eqn
p_2(k)=\cor{\left(\f{1}{2^k} \sum_i l_i+r_i\right)^2} &=&
\f{1}{4}\Biggl\langle \left(\f{1}{2^{k-1}}\sum_i l_i\right)^2+
\left(\f{1}{2^{k-1}}\sum_i r_i\right)^2+ \nonumber\\
&&+2\left(\f{1}{2^{k-1}}\sum_i l_i\right)
\left(\f{1}{2^{k-1}}\sum_i r_i\right)
\Biggr\rangle.
\eqnx
Using the fact that the left and right bins are independent one gets
the recurrence relation:
\eq
p_2(k)=\f{1}{2}\underbrace{(p_a a^2+p_b b^2)}_{\textstyle d_2}
p_2(k-1)+\f{1}{2}.
\eqx
This can be solved together with the initial data (\ref{e.init}), to
yield a closed form solution:
\eq
\label{e.ptwo}
p_2(k)=\left(\f{d_2}{2}\right)^k \cdot \f{1-d_2}{2-d_2}+\f{1}{2-d_2}.
\eqx

In general one can obtain the recurrence relation for general $q$ in
exactly the same way:
\eq
p_q(k)=\f{1}{2^q}\sum_{i=0}^q
\left(\begin{array}{c} q\\i \end{array}\right) d_i d_{q-i} \cdot
p_i(k-1)p_{q-i}(k-1)
\label{pk}
\eqx
where
\eq
d_i=p_a a^i+p_b b^i.
\eqx

\section{Discussion}


We have performed numerical simulations of the $\alpha$-model in order to test
the improved single data analysis in practice.
In Fig.\ 1a-d the histograms of the corrected (with the shift
(\ref{pk}) taken into
account) and standard (without the correction (\ref{pk})) values of
intermittency exponents $\varphi_{2},\varphi_{3}$
are plotted for 90000 generated cascades of 5 and 10 steps. The
peaks with the correction included are significantly closer to the
theoretical value.
The dispersion of the distribution estimated directly from the observed peak,
for the "corrected" histogram is smaller than the dispersion of the
"standard" one.
It decreases with the number of cascade steps. The numerical values of
"corrected" and "standard" dispersion as a function of the cascade length
are presented in Tables 1, 2 for 2 different sets of cascade
parameters.
The corrected dispersion is relatively small, and it allows
to distinguish between the cascades with different parameters (Figs.1a-d).

The influence of the correction (\ref{pk}) on the value of the intermittency
exponents obtained from averaging over the ensemble of events (`center
of mass' of the histogram) was also
investigated. The results are presented in Tables 3,4 for 2 different sets
of cascade parameters. The estimation of intermittency exponents
for the corrected case is much better than for the standard one.


In the preceding, the formula for the correction (see
e.g. (\ref{e.ptwo})) depends on the values
of the parameters $a$, $b$ of the $\al$-model. In practice, however,
one would like to implement some sort of model independent correction.
A possible way of doing this is to use the fact that the corrections
$\log p_2(i)$ and $\log p_3(i)$ seem to change most dramatically in
the first few steps of the reconstruction procedure (near the `end' of
the cascade). After that they seem to stabilize at some constant
value. This would suggest using just the reconstructed moments near
the beginning of the cascade in the fit (\ref{e.fit}). In practice,
however, this might perhaps suffer from low statistics and large
fluctuations.


An alternative procedure would be to first determine the parameters
$a$ and $b$ using the standard (uncorrected) method, and then
substitute those parameters into (\ref{pk}) and use the improved
analysis to obtain a better approximation of the exponents. One could
repeat this until the result no longer changed.

\section{Conclusions}

Our conclusions can be summarized as follows~:\\

(a) the value of intermittency exponent estimated from the maximum of
"corrected" histogram moves closer to the theoretical value,\\

(b) the dispersion of the distribution estimated directly from the
observed peak for the "corrected" histogram is smaller than the
dispersion of the "standard" one,

(c) the corrected value of intermittency exponent obtained after averaging over
the sample of events estimates the theoretical value better than in the
standard case,

(d) a possible procedure of improving the analysis without
the knowledge of $\al$-model parameters is proposed.

\section*{Acknowledgments}

We would like to thank A. Bia{\l}as for discussions and suggestions. This work
was supported by Polish Government grant Project (KBN) grant
2P03B00814 and 2P03B04214.
RAJ and BZ were supported by the Foundation for Polish Science (FNP).

\newpage
\noindent
{\bf Figure captions}
Histograms of the intermittency exponents $\phi_2$ (left
column) and  $\phi_3$ (right column) simulated for the set of
parameters $a=0.8$, $b=1.1$ (upper row) and $a=0.5$, $b=1.5$ (lower
row) in 90000 events for 5 and 10 cascade steps. 
The wider curves correspond to 5 stages of the cascade. `Solid'
curves represent the histograms with the correction (\ref{pk}) taken
into account.
\newpage
\begin{table}[hbpt]
\noindent
Table 1.\ Standard and corrected intermittency exponents (determined
from the position of the maximum of the histograms) and their dispersions
(errors) for $a=0.8,\, b=1.1$ and $n=5,\ldots,10$ cascade
steps. Theoretical values 
for intermittency exponents are $\varphi_{2,theor}=2.85\times 10^{-2}$
and $\varphi_{3, theor}=8.13\times 10^{-2}$. 

\begin{center}
\begin{tabular}{|l|c|c|c|c|c|c|}
\hline \hline
$\varphi_{i}=10^{-2}\times\!\!$& 5           &    6        &      7      &    8        &    9        &   10 \\
\hline
$\varphi_{2}$              &$1.90\pm0.89$&$2.00\pm0.82$&$2.26\pm0.75$&$2.49\pm0.66$&$2.52\pm0.59$&$2.46\pm0.52$\\
\hline
$\varphi_{2,corr}$         &$2.66\pm0.75$&$2.79\pm0.66$&$2.66\pm0.56$&$2.85\pm0.46$&$2.85\pm0.43$&$2.85\pm0.36$\\
\hline
$\varphi_{3}$              &$5.66\pm2.46$&$5.66\pm2.46$&$6.48\pm2.05$&$6.64\pm1.72$&$6.81\pm1.72$&$6.72\pm1.49$\\
\hline
$\varphi_{3,corr}$         &$7.79\pm2.13$&$7.63\pm1.81$&$7.62\pm1.64$&$8.00\pm1.40$&$8.12\pm1.15$&$8.12\pm0.98$\\
\hline \hline
\end{tabular}
\end{center}
\end{table}
\begin{table}[hbpt]
\noindent
Table 2.\ Intermittency exponents (determined from the position of the
maximum of the histograms) and their dispersions (errors) for $a=0.5,\, b=1.5$ 
and $n=5,\ldots,10$ cascade steps. Theoretical values
for intermittency exponents are $\varphi_{2,theor}=3.22\times 10^{-1}$
and $\varphi_{3, theor}=8.07\times 10^{-1}$.
\begin{center}
\begin{tabular}{|l|c|c|c|c|c|c|c|}
\hline \hline
$\varphi_{i}=10^{-1}\times\!\!$     &5            &    6        &    7        &    8        &    9        &   10       \\
\hline
$\varphi_{2}$                   &$2.00\pm1.00$&$2.09\pm0.92$&$2.43\pm0.74$&$2..61\pm0.70$&$2.44\pm0.74$&$2.26\pm0.70$\\
\hline
$\varphi_{2,corr}$              &$3.13\pm0.79$&$3.13\pm0.74$&$2.96\pm0.61$&$2.87\pm0.57$&$3.05\pm0.52$&$3.00\pm0.48$\\
\hline
$\varphi_{3}$                   &$4.52\pm2.23$&$5.31\pm2.23$&$5.83\pm2.02$&$5.90\pm1.83$&$6.03\pm1.83$&$5.70\pm1.70$\\
\hline
$\varphi_{3,corr}$              &$7.53\pm1.90$&$7.93\pm1.83$&$7.66\pm1.57$&$7.53\pm1.31$&$7.66\pm1.31$&$7.66\pm1.18$\\
\hline \hline
\end{tabular}
\end{center}
\end{table}
\newpage
\begin{table}[hbpt]
\noindent
Table 3.\ Standard and corrected intermittency exponents and their dispersions
(errors) for $a=0.8,\, b=1.1$ and $n=5,\ldots,10$ cascade steps obtained after averaging
over the sample of 90000 events.Theoretical values
for intermittency exponents are $\varphi_{2,theor}=2.85\times 10^{-2}$
and $\varphi_{3, theor}=8.13\times 10^{-2}$.

\begin{center}
\begin{tabular}{|l|c|c|c|c|c|c|c|}
\hline \hline
$\varphi_{i}=10^{-2}\times\!\!$     & 5         &    6       &      7    &    8      &    9       &   10 \\
\hline
$\varphi_{2}$                   &$2.16\pm4.34$&$2.45\pm1.51$&$2.57\pm0.74$&$2.63\pm0.66$&$2.68\pm0.60$&$2.72\pm0.54$\\
\hline
$\varphi_{2,corr}$              &$2.90\pm0.70$&$2.89\pm0.58$&$2.88\pm0.49$&$2.88\pm0.42$&$2.87\pm0.36$&$2.87\pm0.31$\\
\hline
$\varphi_{3}$                   &$6.54\pm4.80$&$7.03\pm2.62$&$7.33\pm2.11$&$7.53\pm1.90$&$7.68\pm1.73$&$7.77\pm1.58$\\
\hline
$\varphi_{3,corr}$              &$8.32\pm2.00$&$8.29\pm1.67$&$8.25\pm1.40$&$8.23\pm1.18$&$8.22\pm1.02$&$8.20\pm0.89$\\
\hline \hline
\end{tabular}
\end{center}
\end{table}
\begin{table}[hbpt]
\noindent
Table 4.\ Intermittency exponents and their dispersions (errors) for $a=0.5,\, b=1.5$
and $n=5,\ldots,10$ cascade steps obtained after averaging
over the sample of 90000 events. Theoretical values
for intermittency exponents are $\varphi_{2,theor}=3.22\times 10^{-1}$
and $\varphi_{3, theor}=8.07\times 10^{-1}$.

\begin{center}
\begin{tabular}{|l|c|c|c|c|c|c|c|}
\hline \hline
$\varphi_{i}=10^{-1}\times\!\!$     & 5         &    6       &      7    &    8      &    9       &   10 \\
\hline
$\varphi_{2}$                   &$2.33\pm1.2$&$2.50\pm0.95$&$2.62\pm0.82$&$2.69\pm0.76$&$2.77\pm0.74$&$2.81\pm0.72$\\
\hline
$\varphi_{2,corr}$              &$3.20\pm0.70$&$3.17\pm0.63$&$3.15\pm0.57$&$3.15\pm0.52$&$3.14\pm0.47$&$3.14\pm0.43$\\
\hline
$\varphi_{3}$                   &$5.78\pm2.36$&$6.13\pm2.06$&$6.38\pm1.90$&$6.55\pm1.81$&$6.71\pm1.78$&$6.81\pm1.75$\\
\hline
$\varphi_{3,corr}$              &$8.16\pm1.71$&$8.05\pm1.51$&$7.96\pm1.36$&$7.90\pm1.24$&$7.86\pm1.13$&$7.84\pm1.06$\\
\hline \hline
\end{tabular}
\end{center}
\end{table}
\newpage
%
%
\noindent
\begin{figure}[t]
\hfill a)\epsfig{width=6cm, file=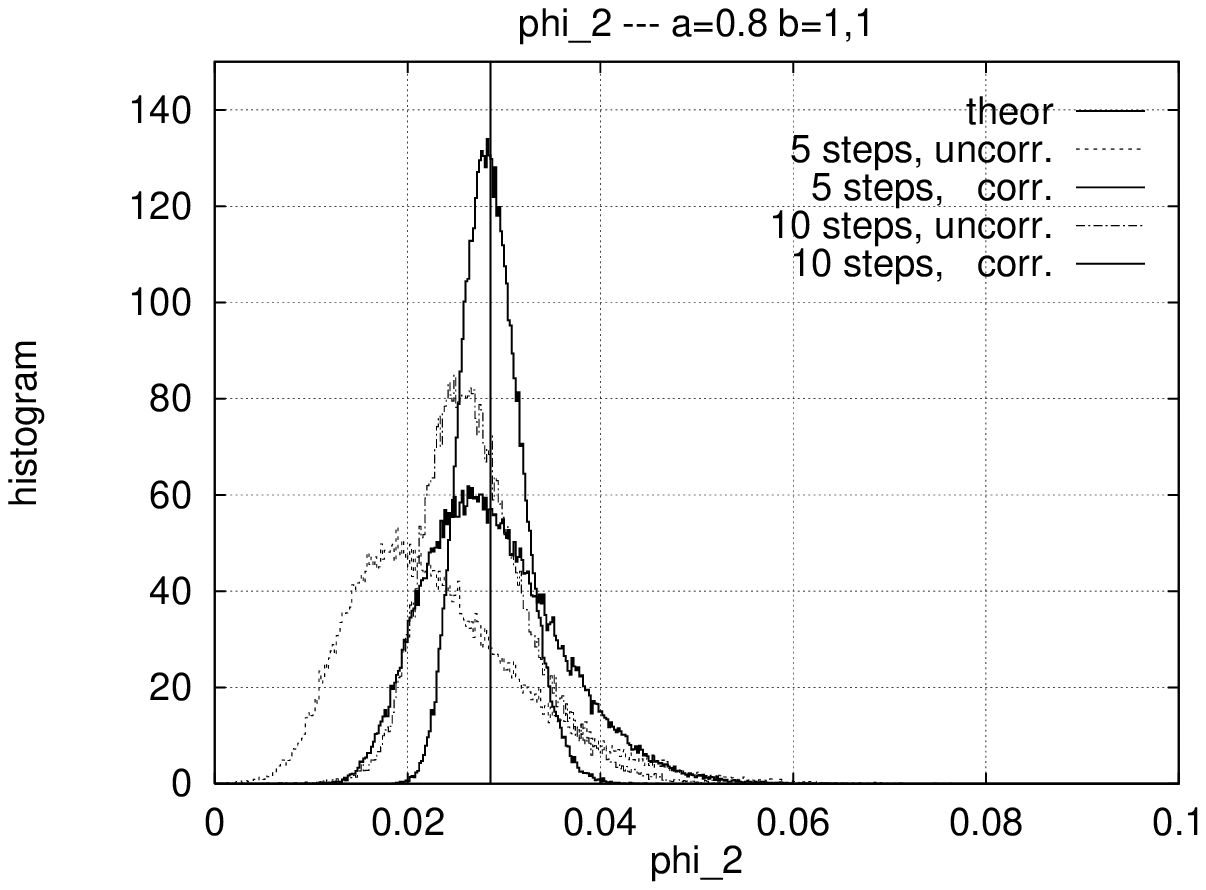}\hfill
b)\epsfig{width=6cm, file=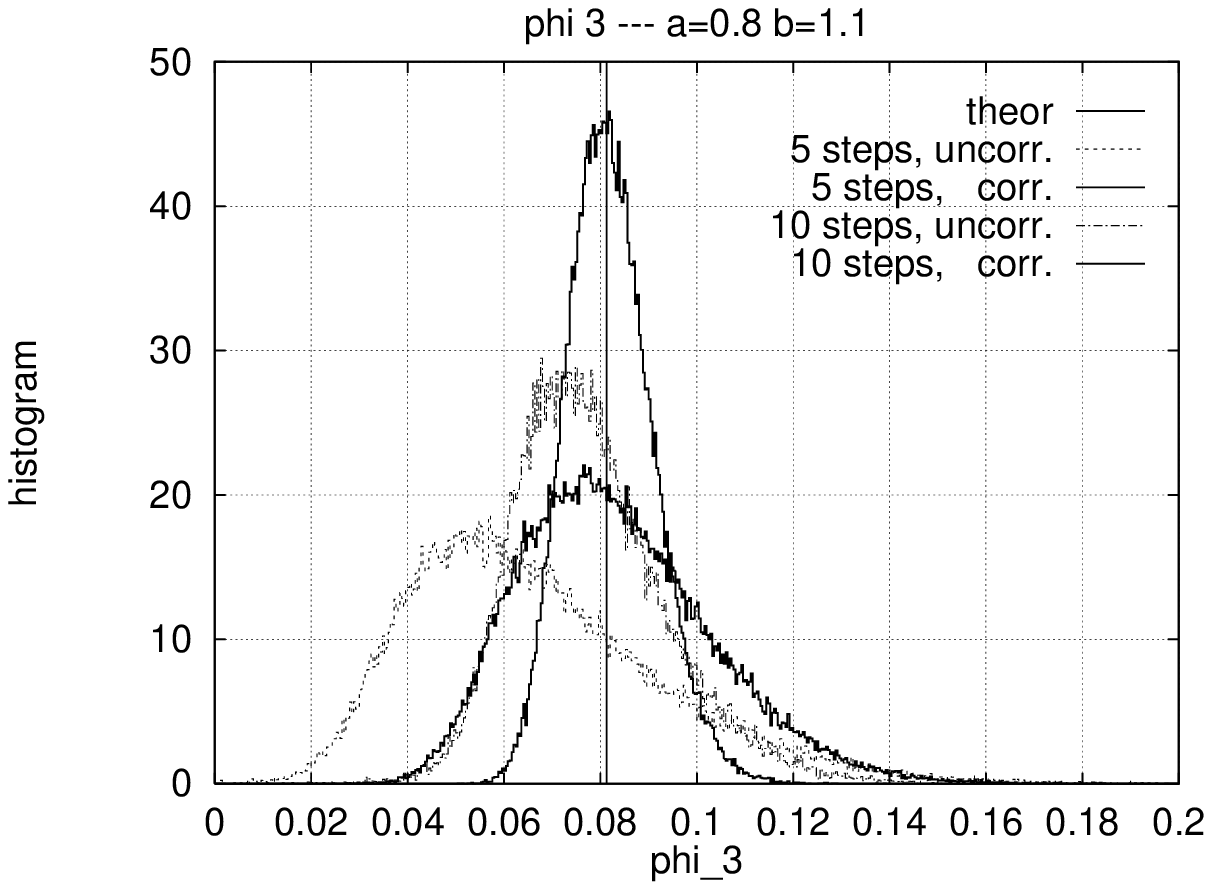}\hfill\mbox{}\vspace{0.5cm}\\
\mbox{}\hfill c)\epsfig{width=6cm, file=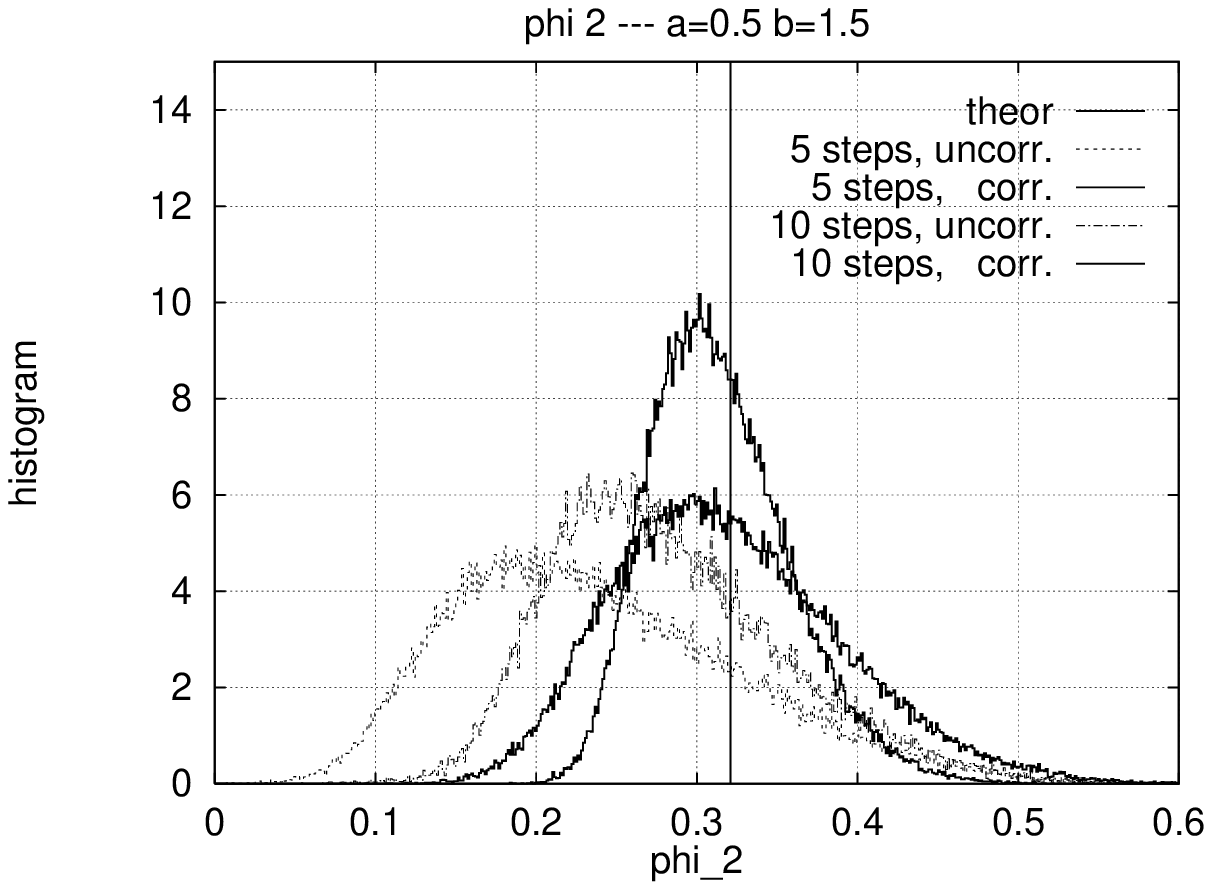}\hfill
d)\epsfig{width=6cm, file=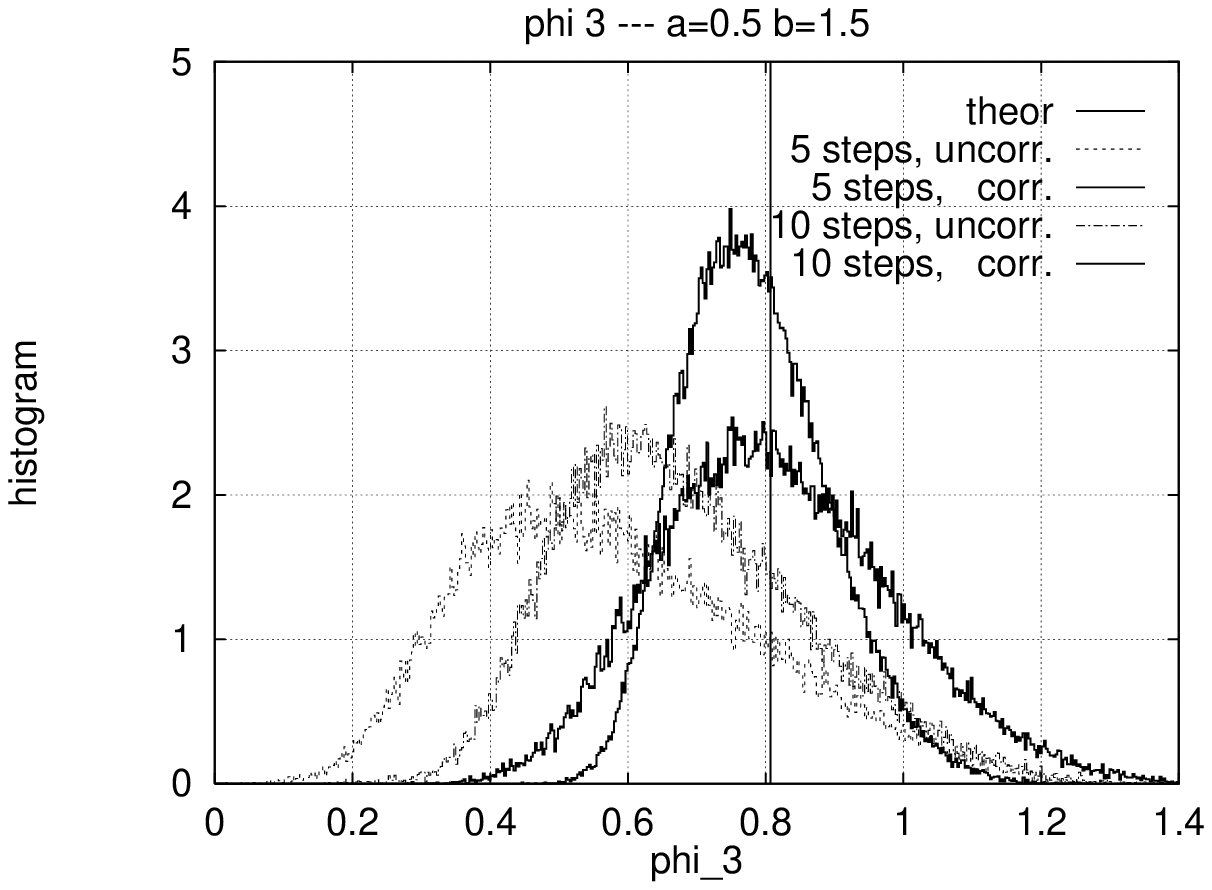}\hfill\mbox{}\\
\caption{a-d}
\label{fig12}
\end{figure}

\end{document}